\documentstyle[preprint,aps,graphicx]{revtex}

\tightenlines

\begin{document}
\draft
\title{
High-precision calculations of van der Waals coefficients
for heteronuclear alkali-metal dimers
}

\author{A. Derevianko\thanks{Permanent Address: Department of Physics, University of Nevada, Reno, Nevada 89557}, J. F. Babb, and A. Dalgarno}
\address{
Institute for Theoretical Atomic and Molecular Physics\\
Harvard-Smithsonian Center for Astrophysics,
Cambridge, Massachusetts 02138}

\date{\today}
\maketitle

\begin{abstract}
Van der Waals coefficients for the heteronuclear alkali-metal dimers
of Li, Na, K, Rb, Cs, and Fr are calculated using relativistic 
{\em ab initio} methods augmented by high-precision experimental data.
We argue that the uncertainties
in the coefficients are unlikely to exceed
about 1\%. 
\end{abstract}


Considerable attention has been given to the determination of the
coefficients of the leading term of the van der Waals attractions of
two alkali metal atoms because of their importance in the simulation,
prediction, and interpretation of experiments on cold atom collisions,
photoassociation and fluorescence
spectroscopy~\cite{BoeTsaGar97,RobClaBur98,WilTieJul99,Boh00,AmiVer00,LeoWilJul00}.
There is strong interest in heteronuclear molecules formed by pairs of
different alkali metal atoms. Experiments have been carried out on
trap loss in mixtures of Na with K~\cite{SanNusMar95,SanNusAnt99},
Rb~\cite{TelMarMun99,YouEjnSha00}, and Cs~\cite{ShaChaBig99} and on
optical collisions ~\cite{ShaChaBig99b} in a Na-Cs mixture and on
molecular formation~\cite{ShaChaBig99a}. The mixtures of
magnetically-trapped alkali metal atoms Na-Cs and Na-K have been
proposed~\cite{EjnRudBig97} as a means to search for evidence of an
electric dipole moment to test for violation of parity and time
reversal symmetry.  We extend here previous studies~\cite{DerJohSaf99}
of the van der Waals coefficient between pairs of identical ground
state alkali metal atoms to unlike ground state atoms.

The leading team of the van der Waals interaction is given at an atom
separation $R$ by~\cite{Goo73,Dal67},
\begin{equation}
\label{Eqn_disp}
  V^{AB} (R) = -\frac{C^{AB}_6}{R^6},
\end{equation}
where
$C_6^{AB}$ is
the van der Waals coefficient.  We use atomic units throughout.

The van der Waals coefficient 
may be expressed as
\begin{equation}
 C_6^{AB} = \frac{2}{3} \sum_{st}
 \frac{ |\langle v_A |D_A | s_A\rangle|^2 |\langle v_B |D_B | t_B\rangle|^2  }
 {(E_s^A -E_v^A) +(E_t^B -E_v^B) } \, , \label{Eqn_C6_direct}
\end{equation}
where $|v_{A}\rangle$ is the ground state atomic wave function of atom
A with energy $E_v^{A}$, and similarly for atom B, and $| s_A \rangle$
and $|t_B\rangle$ represent complete sets of intermediate atomic
states with, respectively, energies $E_s^A$ and $E_t^B$. The
electric dipole operators are 
$D_{A}=\sum_{i=1}^{N_{A}} {\mathbf r}_i^A$, where ${\mathbf r}_i^A$
is the position vector of electron $i$ measured
from nucleus A, $N_{A}$ is
the total number of atomic electrons for atom A,
and similarly for atom B.

At this point the two-center molecular-structure problem is
reduced to the determination of {\em atomic} matrix elements and energies.
The dependence on one-center atomic
properties becomes  explicit when Eq.~(\ref{Eqn_C6_direct})
is cast into the  Casimir-Polder form
\begin{equation}
 C^{AB}_6 = \frac{3}{\pi} \int_0^\infty \alpha_A(i\omega) \alpha_B(i\omega)  
d \omega \, , \label{Eqn_C6_polariz}
\end{equation}
where $\alpha_{A}(i \omega )$ is the dynamic polarizability of
imaginary argument for atom A given by
\begin{equation}
\alpha_A ( i \omega ) = \frac{2}{3} \sum_s
\frac{\left( E^A_s-E^A_v \right) |\langle v^A | {\mathbf D}_A | s^A\rangle|^2}
{ \left( E^A_s-E^A_v \right)^2 +\omega^2 } \, , \label{Eqn_alpha}
\end{equation}
and $\alpha(\omega=0)$ is the  ground-state static dipole 
polarizability. In the limit of infinite 
frequency the function $\alpha_A (i\omega )$ satisfies
\begin{equation}
\alpha_A ( i \omega  ) \rightarrow \frac{N_A}{\omega^2} \, , \label{Eqn_TRK}
\end{equation}
as a consequence of the nonrelativistic Thomas-Reiche-Kuhn  sum rule.

Modern all-order many-body methods are capable of predicting
electric-dipole matrix elements for principal transitions and energies
in alkali-metals to within errors approaching 0.1\%~\cite{SafJohDer99}.
Many-body methods augmented by high-precision experimental
data for principal transitions, similar to those employed in PNC
calculations~\cite{BluJohSap90}, have led to a high-precision evaluation of dynamic
dipole polarizabilities for alkali-metal atoms~\cite{DerJohSaf99}. 
The values of $C_6$  previously calculated for
{\em homonuclear} dimers~\cite{DerJohSaf99} are in  excellent agreement with
analyses of cold-atom scattering of  Na~\cite{vanAbeVer99},
Rb~\cite{RobClaBur98}, and Cs~\cite{LeoWilJul00,ChiVulKer00}.
Here we employ the same methods
to compute the van der Waals coefficients for heteronuclear  alkali-metal 
dimers. 

Precise nonrelativistic variational calculations of $C_6$ for Li${}_2$
have been carried out~\cite{YanBabDal96}.  They provide a critical
test of our procedures.  We separate the dynamic polarizability into
valence and core contributions, which correspond respectively to
valence-electron and core-electron excited intermediate states in the
sum, Eq.~(\ref{Eqn_alpha}). In our calculations for Li we employ
high-precision experimental values for the principal transition
$2s-2p_J$, all-order many-body data and experimental energies for
$3p_{J}$ and $4p_{J}$ intermediate states, and Dirac-Hartree-Fock
values for higher valence-electron excitations.  The high-precision
all-order calculations were performed using the relativistic
linearized coupled-cluster method truncated at single and double
excitations from a reference
determinant~\cite{SafJohDer99,BluJohLiu89}.  Contributions of
valence-excited states above $4p_J$ were obtained by a direct
summation over a relativistic B-spline basis set~\cite{JohBluSap88}
obtained in the ``frozen-core'' ($V^{N-1}$) Dirac-Hartree-Fock
potential.  Core excitations were treated with a highly-accurate
relativistic configuration-interaction method applied to the
two-electron Li$^{+}$ ion.  For the heavier
alkali-metals~\cite{DerJohSaf99} the  random-phase
approximation~\cite{AmuChe75} was used to calculate this contribution.

The principal transition $2s-2p_{J}$
accounts for 99\% of the static polarizability and 96\% of the
Li$_2$ dispersion coefficient.  In  accurate experiments 
McAlexander {\em et al.}~\cite{McAAbrHul96} reported a 
lifetime of the $2p$ state of 27.102(9) ns (an accuracy of 0.03\%) 
and Martin
{\em et al.}~\cite{MarAubBac97} reported 27.13(2) ns. In our
calculations we employ the more precise value from
Ref.~\cite{McAAbrHul96};  in the subsequent error
analysis we arbitrarily assigned an error bar of twice the
quoted value of
Ref.~\cite{McAAbrHul96}, so that the two experiments are
consistent.

The dynamic core polarizability of Li was obtained in the framework of
the relativistic configuration-interaction (CI) method for helium-like
systems.  This CI setup is described by Johnson and
Cheng~\cite{JohChe96}, who  used it to calculate precise
relativistic static dipole polarizabilities.  We extended their method
to calculate the {\em dynamic} polarizability $\alpha(i \omega)$ for
two-electron systems.  The numerical accuracy was monitored by
comparison with results of Ref.~\cite{JohChe96} for the static
polarizability of Li$^+$ and with the sum rule, Eq.~(\ref{Eqn_TRK}),
in the limit of large frequencies. Core-excited states contribute only
0.5\% to $C_6$ and 0.1\% to $\alpha(0)$ for Li. Their contribution
becomes much larger for heavier alkali metals.

\begin{figure}[h]
\begin{center}
\includegraphics*[scale=0.5]{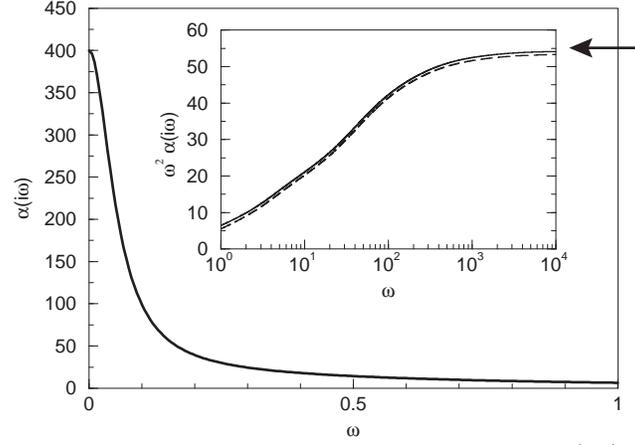}
\caption{
\label{Fig_alpha} 
The dependence of the dynamic dipole polarizability 
$\alpha(i \omega)$ with frequency $\omega$ for Cs.
The inset illustrates the behavior
of the quantity $\omega^2 \alpha(i \omega)$ at asymptotically
large $\omega$, where the dashed line represents the contribution
of the core-excited states to the total $\omega^2 \alpha(i \omega)$
(solid line) and
the arrow marks the non-relativistic
limit $N=55$ following from the sum rule, Eq.~(\protect\ref{Eqn_TRK}).
All quantities are in atomic units.
}
\end{center}
\end{figure}

We calculated static and dynamic polarizabilities and used quadrature,
Eq.~(\ref{Eqn_C6_polariz}), to obtain the dispersion coefficient. The
results are $C_6=1390$ and $\alpha(0) = 164.0$.  There are two
major sources of uncertainties in the final value of $C_6$ ---
experimental error in the dipole matrix elements of the principal
transition, and  theoretical error related to higher
valence-electron excitations. The former results in a uncertainty 
of 0.12\%, 
and the latter much less.  
The  result $C_6 = 1390(2)$ is in  good agreement with the
{\em nonrelativistic} variational result of Yan {\em et
al.}~\cite{YanBabDal96}, $C_6=1393.39$. 
The slight discrepancy between the two values may arise
because 
in our formulation, the correlations
of core-excited states with the valence electron were disregarded 
as were intermediate states  containing
simultaneous excitation of the valence electron with one or both
core electrons.
On the other hand, 
Ref.~\cite{YanBabDal96} did not account for relativistic corrections. 
Relativistic contractions  lead to a smaller value of $C_6$ and
to  better agreement between the present result and
that of Ref.~\cite{YanBabDal96}. 
Similar error analysis for the static polarizability of Li leads to 
$\alpha(0) = 164.0(1)$, which agrees with the numerically precise 
nonrelativistic result of 164.111~\cite{YanBabDal96}. 
An extensive comparison with other published data for the values of 
$\alpha(0)$ and $C_6$ for lithium is given  in Ref.~\cite{YanBabDal96}.
For the heavier alkali metal atoms we followed the
procedures of Ref.~\cite{DerJohSaf99} to calculate
$\alpha (i\omega)$. The results for Cs are illustrated in
Fig.~\ref{Fig_alpha}. They indicate that while most of
the contribution to $C_6$ comes from the resonant transition
at $\omega \sim 0.05$ a.u. 
the core excitations are significant.

%
%

{\em Results and Conclusions\/}---We evaluated the dispersion
coefficients for various heteronuclear alkali-metal dimers with the
quadrature Eq.~(\ref{Eqn_C6_polariz}).  The calculated values are
presented in Table~\ref{Tab_crossC6}.  Most of the contributions to
$C_6^{AB}$ come from the principal transitions of each atom.  An
analysis of the dispersion coefficient of unlike atoms yields the
approximate formula
\begin{equation}
\label{c6-approx}
   C_6^{AB} \approx \frac{1}{2} \sqrt{ C_6^{AA} C_6^{BB} } 
   \frac{\Delta E_A + \Delta E_B}{\sqrt{\Delta E_A \Delta E_B }} ,
\end{equation}
where the energy separations of the principal transitions
are designated  as 
$\Delta E_{A}$ and $\Delta E_{B}$.
Eq.~(\ref{c6-approx}) combined with the high-accuracy values of $C_6$
for homonuclear dimers~\cite{DerJohSaf99} gives accurate
approximations to our results based on Eq.~(\ref{Eqn_C6_polariz}).
For example, Eq.~(\ref{c6-approx}) overestimates our accurate value from
Table~\ref{Tab_crossC6} for Li-Na by 0.4\% and for Cs-Li by 2\%.
We may use Eq.~(\ref{c6-approx}) to estimate the uncertainties 
${\delta C_6^{AB}}$ 
in the
heteronuclear cases from the uncertainties ${\delta C_6^{AA}}$
and ${\delta C_6^{BB}}$  in the homonuclear dispersion
coefficients,
\[
   \frac{\delta C_6^{AB}}{C_6^{AB}} \approx \frac{1}{2} 
   \left[ \left( \frac{\delta C_6^{AA}}{C_6^{AA}} \right)^2 +
   \left( \frac{\delta C_6^{BB}}{C_6^{BB}} \right)^2
   \right]^{1/2} \,.
\]
The accuracy of $C_6$ for homonuclear dimers was assessed
in Ref.~\cite{DerJohSaf99} and
a detailed discussion for the Rb dimer is given in Ref.~\cite{SafJoh00}.
Analyzing the error in this manner using the
quoted coefficients
and their uncertainties from Ref.~\cite{DerJohSaf99} we find that
most of the dispersion coefficients reported here have an estimated
uncertainty below 1\%. The corresponding values
are given in parentheses in Table~\ref{Tab_crossC6}.

\begin{figure}[h]
\begin{center}
\includegraphics*[scale=0.65]{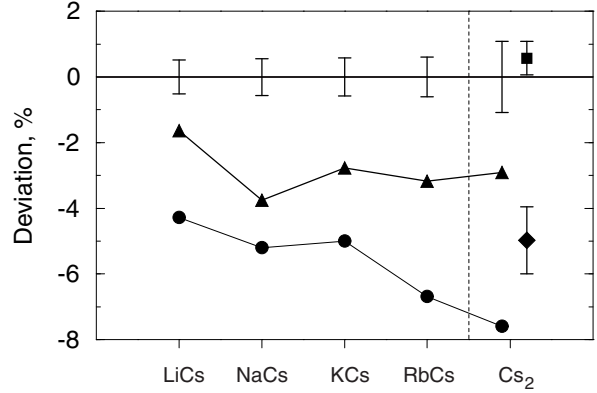}
\caption{
\label{Fig_C6CsX} 
Percentage deviation  of results of  recent 
calculations~\protect\cite{MarSad99,PatTan97} from our values for 
van der Waals coefficients $C_6$ for Cs-Li, Cs-Na, Cs-K,Cs-Rb, and Cs-Cs. 
The values with error bars 
placed along the horizontal line at 0 
correspond to our results with the estimated uncertainties.
Circles represent the results of Ref.~\protect\cite{MarSad99} and triangles the
results of Ref.~\protect\cite{PatTan97}.
For   
Cs-Cs,
to the right of the vertical dotted line,
we  show the  difference between our present prediction,
our earlier 
prediction~\protect\cite{DerJohSaf99}
and the values deduced from cold-collision data in 
Ref.~\protect\cite{LeoWilJul00} (square) and 
Ref.~\protect\cite{DraTolTja00} (diamond).
}
\end{center}
\end{figure}

In Fig.~\ref{Fig_C6CsX} we present for the dispersion coefficients of
the dimers involving Cs a comparison between our calculated values and
the most recent determinations~\cite{LeoWilJul00,DraTolTja00}.  We
give the percentage deviation from our calculations. It is apparent
that the other calculations that employed one-electron model
potentials and accordingly omitted contributions from core-excited
states yield values systematically smaller than ours.

The discrepancies are most significant for Cs${}_2$ where the
number of electrons is greatest.
Fig.~\ref{Fig_C6CsX} also compares the  values 
for the Cs$_2$ dimer with  values deduced from ultracold-collision
data~\cite{LeoWilJul00,DraTolTja00}.  
The agreement of our prediction 6851(74)~\cite{DerJohSaf99} with their  
values for $C_6$ in Cs$_2$ is close. 
Core-excited states contribute  15\%~\cite{DalDav67,DerJohSaf99} 
to the value of the $C_6$ coefficient
for the Cs dimer 
and 
are needed to  fulfill the oscillator strength 
sum rule, Eq.~(\ref{Eqn_TRK}). In the present approach the 
contributions of core-excited states to dynamic polarizabilities
are obtained using the random-phase approximation,
which  nonrelativistically 
satisfies 
the oscillator strength  sum rule exactly~\cite{AmuChe75}.
In the inset of Fig.~\ref{Fig_alpha}, it is  illustrated
that our calculated $\alpha (i\omega)$ approaches $N/\omega^2$
as $\omega$ becomes asymptotically large,
where $N=55$ for Cs.
While the deviation between 
the present calculations and the model potential calculations
are smaller for dimers involving lighter atoms,
an accurate accounting of core-excited states is essential
to achieve high accuracy in dispersion coefficient
calculations for heavy atoms~\cite{DalDav67,MaeKut79,MarSadDal94}.

Few experimental data are available for comparison in the
heteronuclear case, except for NaK.  The results from investigations
of NaK molecular potentials based on spectral
analysis~\cite{RusRosAub00} are compared to our value in
Table~\ref{Tab_mol}.  Our value is smaller than the experimental
values. Earlier theoretical calculations of dispersion coefficients
for NaK have been tabulated and evaluated by Marinescu
and Sadeghpour~\cite{MarSad99} and by Zemke and Stwalley~\cite{ZemStw99}. 
Those values are generally lower than our value  
of $2447(6)$ except for that of Maeder and Kultzelnigg~\cite{MaeKut79}
who give $2443$.


The present study extends the application of modern relativistic
atomic structure methods to calculations of ground state van der Waals
coefficients of Li${}_2$ and of the heteronuclear alkali-metal atoms.
We argue that the uncertainty of the coefficients is unlikely
to exceed 1\%.
Additional experimental data from future cold-collision experiments or
spectroscopy would provide further tests of the present calculations.

This work was supported by the Chemical Sciences, Geosciences and
Biosciences Division of the Office of Basic Energy Sciences, Office of
Science, U.S. Department of Energy and by the National Science
Foundation under grant PHY97-24713. The Institute for
Theoretical Atomic and Molecular Physics is supported
by a grant from the NSF to Harvard University
and the Smithsonian Institution.

%
%
\clearpage
%

\clearpage
%
%
\begin{table}
\caption{ 
Dispersion coefficients $C_6$
and their estimated uncertainties (parentheses)
for alkali-metal atom pairs
in atomic units.
Coefficients for 
Na$_2$, K$_2$, 
Rb$_2$, Cs$_2$, and Fr$_2$ are from Ref.~\protect\cite{DerJohSaf99}.
\label{Tab_crossC6} }
\begin{tabular}{lllllll}
\multicolumn{1}{c}{}&
 \multicolumn{1}{c}{Li}&
 \multicolumn{1}{c}{Na}&
 \multicolumn{1}{c}{K}&
 \multicolumn{1}{c}{Rb}&
 \multicolumn{1}{c}{Cs}&
 \multicolumn{1}{c}{Fr}\\
 Li  &1389(2)&1467(2)&2322(5)&  2545(7) &   3065(16) &  2682(23) \\
 Na  &     & 1556(4)& 2447(6) &  2683(7) &  3227(18)&  2842(24) \\
 K   &     &	    & 3897(15)&  4274(13)&  5159(30)&  4500(39) \\
 Rb  &     &	    &	      &  4691(23)&  5663(34)&  4946(44) \\
 Cs  &     &	    &	      &	         &  6851(74)&  5968(60) \\
 Fr  &     &	    &	      &	         &  	    &  5256(89) \\
\end{tabular}
\end{table}
\clearpage
%
\begin{table}
\caption{ 
Comparision of present theoretical
and experimental values for the dispersion coefficient for NaK.
\label{Tab_mol} }
\begin{tabular}{ll}
\multicolumn{1}{l}{ Reference}&\multicolumn{1}{c}{$C_6$}\\
\hline
This work                                       & 2447(6)  \\
Russier-Antoine {\em et al.,}~\cite{RusRosAub00}& 2519(10)\tablenotemark[1]\\ 
Ishikawa {\em et al.,}~\cite{IshMukTan94}       & 2646(31)\tablenotemark[1]\\
Ross {\em et al.,}~\cite{RosEffDin85}         & 2669.4(20)\tablenotemark[1]\\ 
\end{tabular}
\tablenotetext[1]{Experiment.}
\end{table}


\end{document}